%% file: sample-authordraft.tex
\documentclass[sigconf]{acmart}

\AtBeginDocument{%
  \providecommand\BibTeX{{%
    \normalfont B\kern-0.5em{\scshape i\kern-0.25em b}\kern-0.8em\TeX}}}

\usepackage{multirow}
  
\usepackage{amssymb}
\usepackage{amsmath, amsfonts}
\usepackage{graphicx}
\usepackage{textcomp}
\usepackage{xcolor}
\usepackage{hyperref}
\usepackage{cleveref}
\usepackage{xspace}
\usepackage{paralist}
\usepackage{multirow}
\usepackage{subcaption}
\usepackage{tikz}
\usepackage{courier}
\usepackage{listings} 
\usepackage{balance}
\usepackage{adjustbox}

\usepackage{array}
\usepackage{graphics}
\usepackage{lipsum}

\usepackage{times}
\usepackage{latexsym}

\usepackage[T1]{fontenc}

\usepackage[utf8]{inputenc}

\usepackage{microtype}
\usepackage{xurl}

\newcommand{\todo}[1]{\textcolor{blue}{\bf TODO}{{\footnotesize\bf~#1}}}

\newcommand{\etal}{\emph{et al.}\xspace}
\newcommand{\ie}{\emph{i.e.},\xspace}
\newcommand{\eg}{\emph{e.g.},\xspace}

\begin{document}

\title{Towards Understanding What Code Language Models Learned}

\author{Toufique Ahmed, Dian Yu, Chengxuan Huang, Cathy Wang, Prem Devanbu, Kenji Sagae}
\email{{tfahmed,diayu, cxxhuang, catwang, ptdevanbu, sagae}@ucdavis.edu}
\affiliation{%
  \institution{Department of Computer Science and Department of Linguistics, UC Davis}
  \streetaddress{1, Shields Way}
  \city{Davis}
  \state{CA}
  \country{USA}
 \postcode{95616}
}
\renewcommand{\shortauthors}{Ahmed et al.}

\begin{abstract}
  Pre-trained language models work well for a range of natural language tasks, but some have argued that they are not capable of fully learning meaning or understanding language. To understand the extent to which language models can learn some form of meaning, we investigate their ability to capture semantics of \emph{code} beyond superficial frequency and co-occurrence. In contrast to previous research on probing models for linguistic features, we study pre-trained models in a setting that allows for objective and straightforward evaluation of a model's ability to learn semantics. In this paper, we examine whether such models capture the semantics of code, which is precisely and formally defined. Through experiments involving the manipulation of code fragments, we show that code pre-trained models of code learn a robust representation of the \emph{computational semantics} of code that goes beyond superficial features of form alone.

\end{abstract}

\keywords{PLMs, Semantic Understanding, Robust Learning}



\maketitle

\section{Introduction}
Pre-trained language models (PLMs) such as BERT \cite{devlin-etal-2019-bert}, GPT-3 \cite{brown-etal-2020-langauge}, and PaLM \cite{chowhert-etal-2022-palm} are powerful processors of natural language, proficient at tasks such as question answering and joke explanations \cite{devlin-etal-2019-bert, brown-etal-2020-langauge, chowhert-etal-2022-palm}. 
Their skill is amplified with just a few examples, 
or even natural language instructions without labeled data \cite{ouyang-etal-2022-training}. PLMs also seem quite adept at transfer learning, despite relatively simple pre-training objectives. 
Previous investigations of the linguistic capabilities of PLMs show that they encode syntactic, semantic, and world knowledge \cite{rogers-etal-2020-primer}. However, while some authors suggest that PLMs  understand language and learn meaning,  
others have argued that they do not truly understand the meaning of natural language \cite{bender-koller-2020-climbing}. 

Some of the disconnect between those who tout the ability of PLMs in natural language understanding and those who dispute these claims 
arises from disagreements on what is considered ``understanding'' or ``meaning''. Bender \& Koller~\cite{bender-koller-2020-climbing} argue  that, if meaning is ``the relation between a linguistics form and communicative intent,'' then PLMs may not capture meaning. Still, even though language models lack communicative intent and grounding of language in the world, they do appear to capture semantic relationships that go beyond superficial word frequency and co-occurrence patterns. Setting aside issues of pragmatics, which are crucial to language understanding, we attempt to characterize what PLMs learn in terms of meaning by focusing on their ability to perform a much more restricted form of ``understanding'' at the semantic level. To this end, we examine whether PLMs can capture the \emph{semantics of programs}, which are precise and concrete---thus allowing for straightforward objective experimentation, without requiring theoretical assumptions about natural language semantics.

This question (concerning semantics captured by PLMs for code) arises because, as with natural language tasks, neural \emph{code} language models appear capable on code-related tasks: code summarization,  retrieval, and generation; defect detection; and others \cite{kanade-etal-2020-learning, feng-etal-2020-codebert, guo-etal-2021-graphcode,chen2021evaluating}. 
Indeed, pre-trained neural code  models perform well enough at software-engineering related tasks \cite{lu-etal-2021-codexglue} that some are now incorporated into practical programming tools such as Github CoPilot\footnote{ \url{https://github.blog/2022-03-29-github-copilot-now-available-for-visual-studio-2022/}}.
Given their impressive performance, it is natural to wonder how much PLMs understand about code. We examine specifically whether a PLM's apparent ``understanding'' of code is just capturing distributions of lexical and syntactic token frequencies and co-occurrence patterns, or if PLMs are actually capturing distributions at a deeper level,  \emph{viz.,} of the computational semantics of code: thus in some sense they ``know'' the meaning of the code, having learned an abstract representation of semantics beyond just superficial form. 
Our results provide some evidence suggesting that they are. This is noteworthy, 
especially in light of some recent results suggesting that language models respond
to ``knowledge probes'' in ways very sensitive
to changes in form~\cite{zhou2023navigating}.

\begin{figure*}[!t]
    \centering
    \includegraphics[scale=0.28]{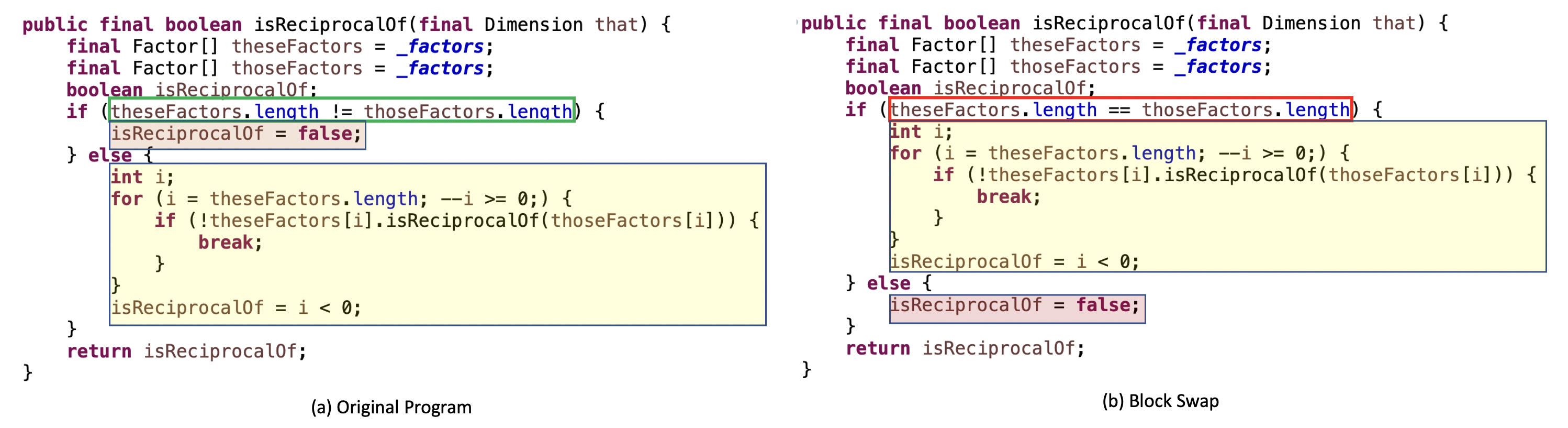}
    \caption{Semantically identical forms after transformation: Block Swap. If the block is swapped, the model should predict ``=='' rather than ``!=''.}
    \label{block}
\end{figure*}

Our approach is largely automated, inspired by the concept of \emph{Metamorphic Testing}. Most of the current PLMs are pre-trained to reconstruct partially-obscured programs, \emph{in the ``original, natural" forms that humans wrote them}. However, because of the formal semantics of programs, there are quite varied syntactic forms of the \emph{very same} computational meaning. If PLMs are robustly learning distributions over computational \emph{semantics}, rather than \emph{lexical} or \emph{syntactic} patterns, then they should also be able to correctly reconstruct \underline{other} forms of programs that \emph{have the same meaning}. 

Programming languages allow one to automatically rewrite code while preserving meaning; if PLMs are modeling  \emph{meaning distributions}, they should be able to reconstruct the rewritten form just as they can the original form. 
Our results suggest that PLMs are learning a sufficiently  robust representation of the \emph{meaning} of code, that is \emph{resistant to at least some manipulations of form}. The methodology we use is to measure how often, and how accurately, a PLM for code can reconstruct (or complete)
code fragments whose form has been manipulated, \emph{while preserving meaning}. Our results show that PLMs learn important aspects of semantics from code corpora without any information about output or execution, indicating that they capture an important part of meaning from form alone.

The meaning is captured with (conditional) low-entropy in the form, as anticipated by Rational Speech Act theory: since the programmer writes code to be read by other programmers, the probability of code given its meaning, denoted as $p(code \mid meaning)$, is low-entropy. This implies that, given a particular meaning, there is often just one ``natural'' way to write the code. Therefore, by learning distributions over code syntax, the model is also learning distributions over meanings. Consider a toy program in Table~\ref{toy}. On the left, we have the original version with the ``=='' operator, and the model is capable enough to predict this token correctly. On the right version, we have swapped the branch of the if statement. What should we expect from the model now? To be syntactically correct, the model could repeat the original ``=='' operator, but instead, it proposes ``!=''. This indicates that the model has learned to resolve artificially-constructed semantic dilemmas, even though it was trained purely in the syntactic form originally written by humans. This is a small program, and we can see that the model understands the semantics of the isEqual function, but can it comprehend the model in a real-world, larger program? This paper attempts to investigate that.

\begin{table}[t]

\resizebox{\columnwidth}{!}{%
\renewcommand{\arraystretch}{1.2}

\centering
\begin{tabular}{ll}
\hline
\multicolumn{1}{c}{Original}                                                                                                                                                                                          & \multicolumn{1}{c}{Transformed}                                                                                                                                                                                       \\ \hline
\begin{tabular}[c]{@{}l@{}}public static boolean isEqual(int a, int b) \{\\         \hspace{.5cm}if (a \textcolor{blue}{==} b) \{\\             \hspace{1cm}return true;   \\         \hspace{.5cm}\} else \{\\             \hspace{1cm}return false; \\         \hspace{.5cm}\}\\     \}\end{tabular} & \begin{tabular}[c]{@{}l@{}}public static boolean isEqual(int a, int b) \{\\         \hspace{.5cm}if (a \textcolor{red}{!=} b) \{\\             \hspace{1cm}return false;   \\         \hspace{.5cm}\} else \{\\             \hspace{1cm}return true; \\         \hspace{.5cm}\}\\     \}\end{tabular}

\\ \hline
\end{tabular}
}
\caption{Example showing two semantically equivalent codes.}
\label{toy}
\vspace{-.8cm}
\end{table}

Previous studies~\cite{karmakar2021pre, troshin2022probing} have attempted to understand pre-trained code models using probing tasks. Probing tasks, also known as diagnostic classifiers, auxiliary classifiers, or decoding, typically use the encoded representations of one system to train another (weak) classifier on a  (probing) task. We emphasize that the \emph{probing}  task is usually different
from the \emph{pre-training} task, and is intended to gauge what the model
has learned during pre-training.
In contrast, we focus solely on the pre-trained (not fine-tuned) model and extract information directly from it. We transform the code to preserve meaning, and ask
the pre-trained model to reconstruct blanked-out bits of the code in the transformed form. We make the following contributions.

\begin{enumerate}
 \item By testing two types of  meaning-preserving transformations (block swap and operand swap), we found that PLMs can accurately predict blanked-out operators in both the original and transformed versions. In most cases, the models were able to predict both versions correctly for the same code snippet.
 \item After applying other meaning-perserving transformations to the code, such as renaming variables and changing the position of conditional statements, we 
 found that the model could still robustly reconstruct blanked-out code correctly. 
 \item We also attempted to identify the relevant contexts that the models use to make predictions. Our experiments indicated that, as expected, the relevant tokens following the pivotal blanked-out point contribute more to the model's semantic understanding than the preceding tokens.
\end{enumerate}

\begin{figure}[t]
    \centering
    \includegraphics[scale=0.35]{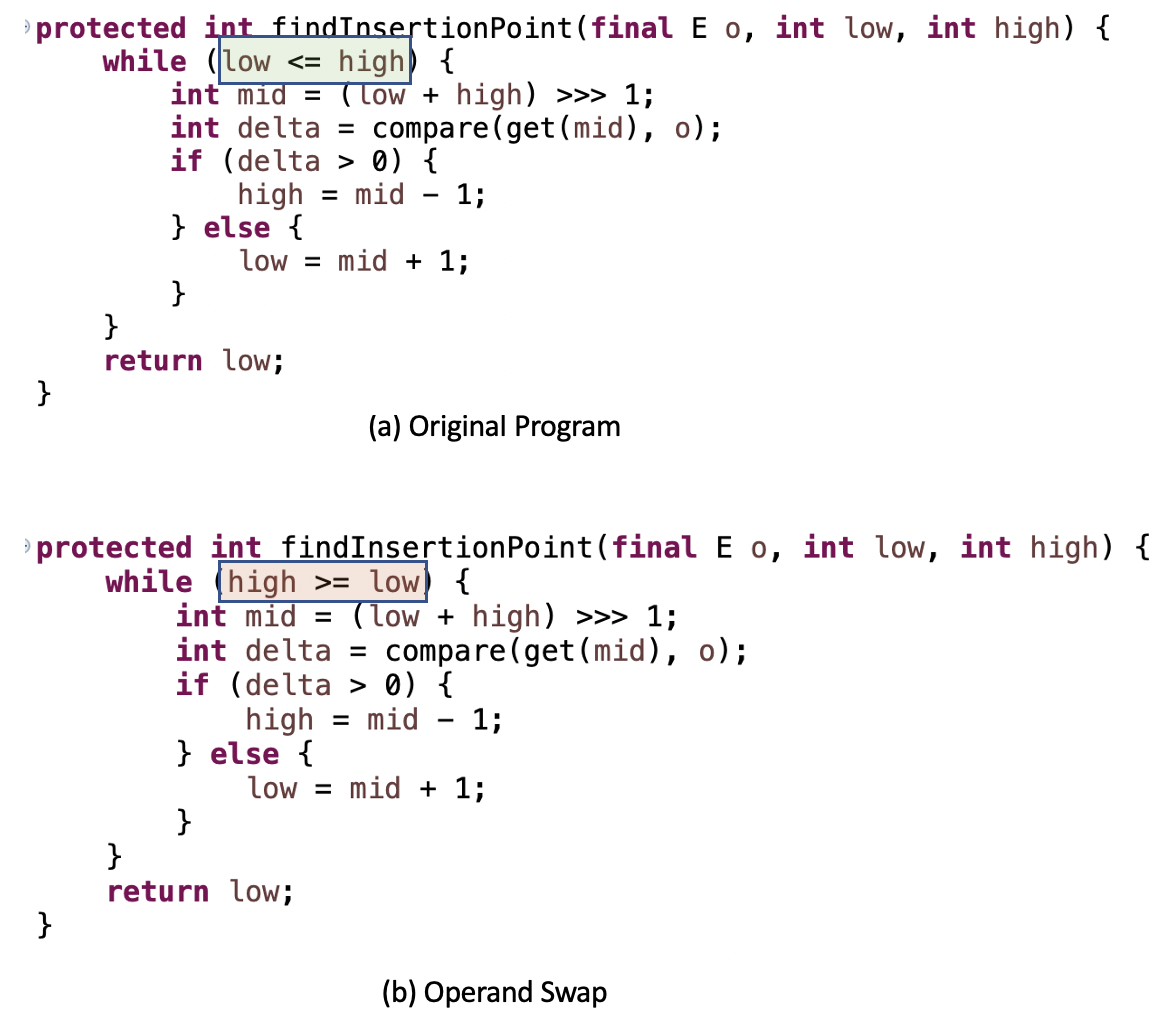}
    \caption{Semantically identical forms after transformation: Operand Swap. If the operand is swapped, the model should predict ``>='' rather than ``<=''.}
    \label{operand}
    \vspace{-.8cm}
\end{figure}

\begin{figure*}[t]
    \centering
    \includegraphics[scale=0.50]{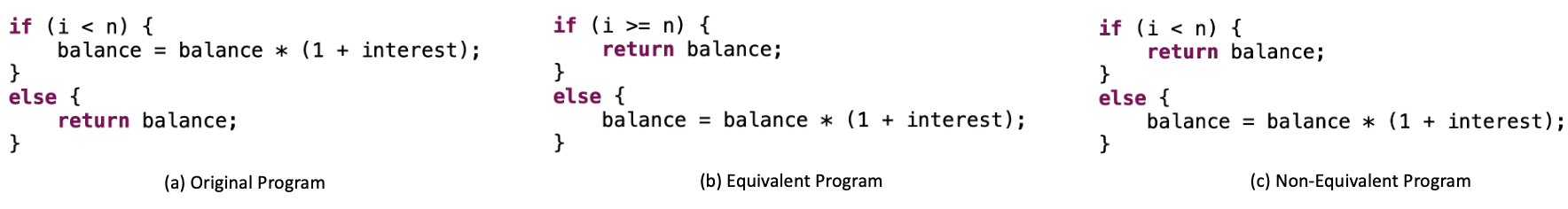}
    \caption{Semantically equivalent and non-equivalent pairs of program snippets. We found that the embedding vector distance between the programs in equivalent pairs is lower than the distance between the programs in nonequivalent pairs, even though the programs in the nonequivalent pairs are more similar to each other than the programs in the equivalent pairs from the superficial perspective of tokens and token sequences.}
    \label{distance}
    \vspace{-.4cm}
\end{figure*}

\section{Related Work}

Language models can perform various tasks such as reading comprehension
\cite{wei-etal-2022-chain, chowhert-etal-2022-palm}. However, model performance is not always robust, so it's unclear to what extent the performance of PLMs indicates an  ``understanding" of the input.
For example, PLMs appear to lack robustness in handling  negation and other changes of  intent in the input \cite{kassner-schutze-2020-negated, ettinger-2020-bert}. Moreover, adversarial triggers \cite{wallace-etal-2019-universal, jin-etal-2020-bert} can distort the model's output, regardless of context. This suggests that PLMs have limited understanding of the ``meaning'' of input texts. 
As in  NLP,  SE researchers are interested in studying what pre-train models learn. Karmakar and Robbes designed four probing tasks (probing for surface-level, syntactic, structural, and semantic information)~\cite{karmakar2021pre}. They used them to show how probes can be used to observe different code properties.
Troshin and Chirkova introduced a new set of more complex probing tasks~\cite{troshin2022probing}. They also consider a more comprehensive range of pre-trained models and investigate different pretraining objectives, model sizes, and the effect of fine-tuning.   Sharma \etal proposes a neuron-level interpretability approach for neural code intelligence models, eliminating redundancy by excluding highly similar or task-irrelevant neurons and evaluating the significance of the remaining ones using probing classifiers~\cite{sharma2023interpreting}. 
Unlike prior works, we avoid  a separate probing classifier and directly interrogate the pre-trained models (via API) to study the robustness of the models' prediction.

Our work also differs from approaches in NLP, which probe the knowledge that natural language models capture: either by correlating model representations with known information \cite{tenney-etal-2018-what, wu-etal-2020-perturbed} or by extracting information from the model \cite{pimentel-etal-2020-information, elazar-etal-2021-amnesic}. Rather, we investigate the degree to which LMs (for code) understand code inputs. Specifically, instead of polluting the input or introducing implicit diagnosing targets \cite{rogers-etal-2020-primer}, we design experiments to show whether 
PLMs simply model the superficial (lexical or syntactic) patterns of code, or whether these models capture a deeper notion of semantics of code.
We note that our work here is based on 
BERT-style models. It's certainly likely that 
very large (generative, auto- regressive) language models (``VLLMs", such as GPT-2~\cite{radford2019language}, GPT-3~\cite{brown-etal-2020-langauge})  also learn the semantics of  programs, perhaps even better the BERT-style models we use. However,
we prefer to use BERT-style models for our experimental work, for two reasons. 
First, BERT-style models are still in wide use. More importantly, it is possible to exercise better experimental control the training/test splits of these models than for the very large Codex-style models (VLLMs). For the smaller models  
(\eg CodeBERT~\cite{feng-etal-2020-codebert} and GraphCodeBERT~\cite{guo-etal-2021-graphcode}) that we use, we have full access to the training and test splits, drawn from CodeSearchNet~\cite{husain2019codesearchnet}. This data has been carefully de-duplicated~\cite{allamanis2019adverse}. 
Thus, for our code transformation experiments, we can study how these models
performed on transformed code drawn from the test split, and interpret the results with greater confidence that this data was not seen during pre-training. 

Prior research has also reported the use of meaning-preserving transforms, to train and/or analyze the behavior of models. For instance, Jain \etal~\cite{jain2020contrastive} utilized an automated source-to-source compiler to generate functionally similar variants of a program, as data, and pre-trained a neural network to pick them out from non-equivalent distractors. Henke \etal~\cite{henke2022semantic} and colleagues employed sequences of parametric, semantics-preserving program transformations in adversarial training to develop models resistant to such adversaries. Chakraborty \etal's Natgen~\cite{chakraborty2022natgen} introduced unnatural forms of code using 6 classes of meaning-preserving transformations, and then pre-trained a model on the task of re-generating the original, more ``natural" programs. Wan \etal~\cite{wan2022they} conducted a thorough structural analysis to interpret pre-trained language models for source code, such as CodeBERT and GraphCodeBERT, from three perspectives, including attention analysis, probing on word embedding, and syntax tree induction. They discovered that integrating the syntax structure of code into the pre-training process may lead to better code representations.

Our goal here is not to further enhance model  performance, but rather to test
\emph{if the model trained in 
the usual, simple, standard, self-supervised pre-training task actually earns a distribution over code meaning}: For instance, BERT models are trained to unmask randomly selected tokens; it's not immediately evident that this very simple training would contribute to the model's resilience to meaning-preserving transformations.

\section{Methodology}
\subsection{Models' Accuracy with Meaning Preserving Transformation}
\label{sec:meaning_preserving_transform}

Unlike natural language, code affords  \emph{feasibly automate-able} meaning-preserving transforms: sometimes, code can be rewritten into lexically and syntactically different forms, while preserving meaning.

\subsubsection{Like Compilers do!}
Compilers, \eg, can change the \emph{form} of code while preserving the computational meaning\footnote{This property
enables effective \emph{metamorphic} testing~\cite{chen2018metamorphic} of compilers.}. In other words, the compiler for a language ${\mathcal L}$ has a built-in, robust conception of the meaning of code in language ${\mathcal L}$.  This robust conception allows compilers (during optimization) to extensively modify the \emph{form} of code, without changing its \emph{meaning}. Thus, here we're asking
the question: \emph{can a PLM, trained on just simple lexical fill-in-the-blanks language modeling tasks, capture semantics, in a similar manner to compilers}? Note that compilers are explicitly programmed to capture programming language semantics, via meaning-preserving transforms, but PLMs are not. 
Here are the two different types of meaning-preserving transformations for our experiments.

\subsubsection{Block Swap}

An example is seen in Figure~\ref{block}: the original java {\tt if} statement is rewritten into the semantically identical form by flipping the {\tt then} and {\tt else} blocks, and also  the condition. The forms are different, but the \emph{semantics} is unchanged. A PLM that learns a robust representation of the code's computational semantics should  perform its primary completion task  correctly, regardless of the lexical or syntactic form of this computation, especially if the computation is very common. In the case of a BERT-like PLM, the primary task is masked language modeling (MLM): \emph{viz.,} fill in a masked token. Thus in the example of the original program in Figure \ref{block}, if the comparison operator  {\tt !=} is masked, a well-trained PLM should guess it correctly if it can determine that one specific computation is preferred over others resulting from alternatives that are equally valid from a syntactic perspective. This preference could well be a reflection of frequency. Now, if the PLM were calculating a \emph{robust} representation of the computation, if the operator were masked in the transformed (block swap), \emph{semantically identical} form, it should guess the {\tt ==} token. 

\subsubsection{Operand Swap}
Figure~\ref{operand} presents another type of meaning-preserving transformation. Unlike block swaps, we did not change the order of the statements; instead, we switch the positions of the operands used for binary operators. If we swap the positions of ``a'' and ``b''  for this expression, $a>b$, we have to update the operator ``>'' to ``<'' to preserve the meaning.  
A variety of such transforms are possible: more generally, assume a token (operator, identifier, etc) $\omega$ in a program $P$. When this program form $P$ is transformed into a semantically equivalent form $P^T$, let us assume that the $\omega$ token in $P$ becomes a  token $\omega^T$ in $P^T$. Given a PLM ${\mathcal M}$,  we check that when $\omega$ is masked out in $P$, ${\mathcal M}$ is able to correctly guess it; in addition, if ${\mathcal M}$ were able to robustly capture the computational semantics of $P$, it should also recover $\omega^T$ in $P^T,$ were that token masked. Successful recovery of $\omega^T$ could not be based primarily on token frequency information, since the $P^T$ is made of the same tokens as $P$ (aside from the masked operator), but in different order. While the frequency of token sequences could be useful information, $P^T$ in general expresses the same meaning in a way that is substantially less frequent than how it is expressed in $P$.

\begin{figure}[t!]
    \centering
    \includegraphics[scale=0.35, trim={0cm 0cm 0cm 0cm}]{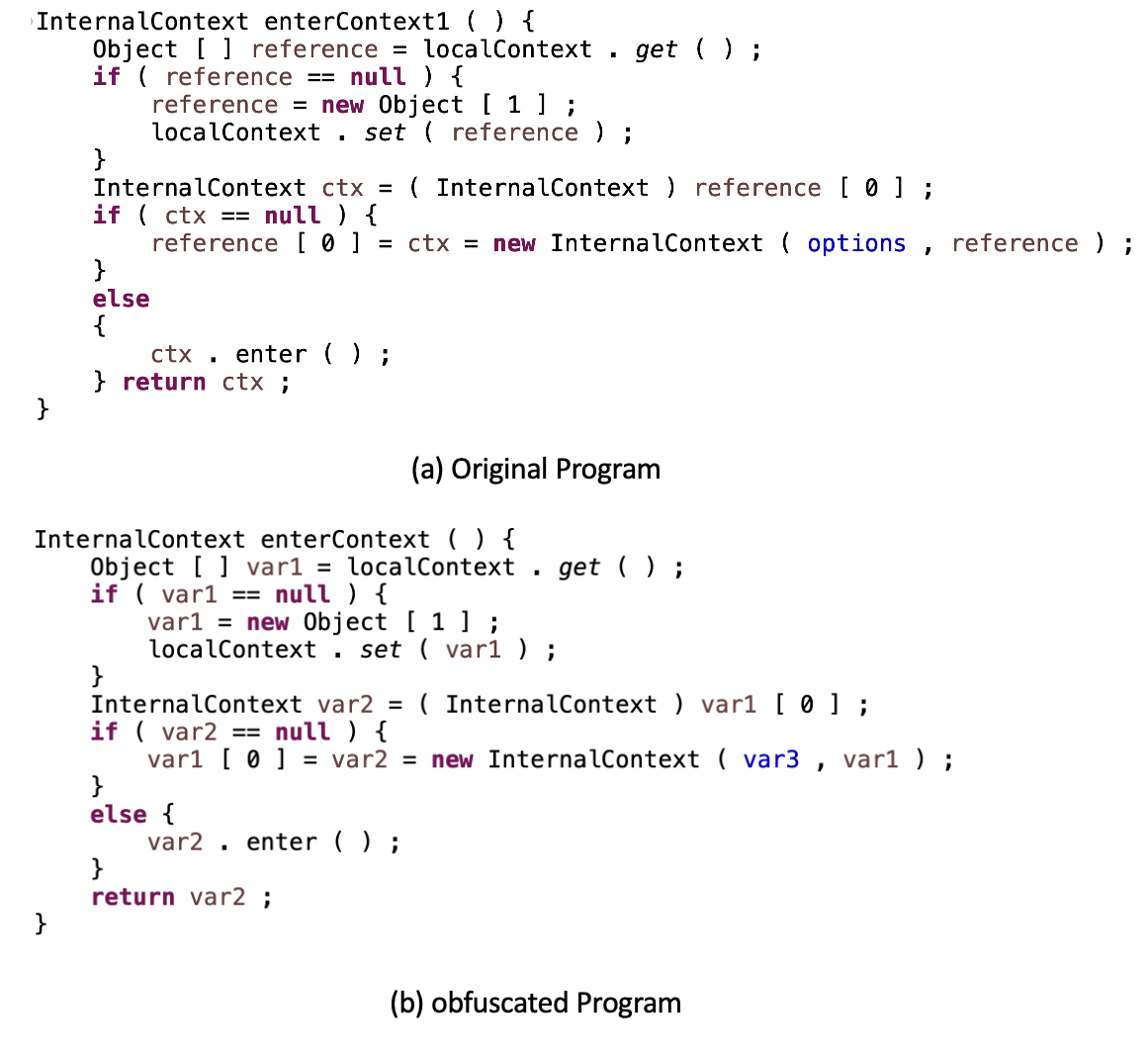}
    \caption{Impact of variable renaming on the performance of GraphCodeBERT model in block swap transformation. Results show that the model's performance degrades by less than 1\%-4\%.}
    \label{var_nam}
    \vspace{-.5cm}
\end{figure}

\subsection{Robustness \& Structure of Semantic Representation}
We now consider some factors affecting how robustly the model appears to be learning the computational meaning of the code. We consider various potential factors that might be relevant, including: the influence of variable naming; the code context that allows the model to learn the meaning; and the effect of refactoring conditions. Finally, we examine the geometry of the internal embedding, to check if similarity in meaning translates to closeness in vector space. 
\subsubsection{Consistent Variable Renaming}
\label{cvr}

Apart from block swap and operand swap, we also evaluated the performance of the GraphCodeBERT model with consistent variable renaming. Renaming the locally declared variables leaves semantics unaffected, while substantially changing lexical form. Figure~\ref{var_nam} presents an example where we replace the {\tt reference} and {\tt ctx} variables with {\tt var1} and {\tt var2} and perform the block swap and operand swap evaluation following the approach we mentioned in Section \ref{sec:meaning_preserving_transform}. We only change the locally-scoped variables, because renaming  other variables may change the semantics on the program; our goal here is to evaluate whether
PLMs can reconstruct programs of the same meaning.  We discuss our findings in Section~\ref{res_var_ren}.

\subsubsection{Context length and direction}
\label{clen}
PLMs use embeddings of context to make
their predictions. If predictions are robust to
changes in form, which part of the context matters in determining the operators? To determine
the operator in an {\tt if} statement, a human developer needs to see the \emph{following code} (\ie code written after the conditional statement) to select the right operator. The code \emph{before} the
{\tt if} statement should have much less influence. 
Is this also true for the model?  
In this study, we  gradually increase the context on both sides (before and after the conditional statements) and  observe the models' output. We compare it with results we get only using the after context. Of course, it's possible that previous tokens may also help maintain the program semantics (\eg method name, identifiers).

\subsubsection{Refactoring of conditional statement}
\label{sec:refactoring}
\begin{figure}[t]
    \centering
    \includegraphics[scale=0.40, trim={0cm 0cm 0cm 0cm}]{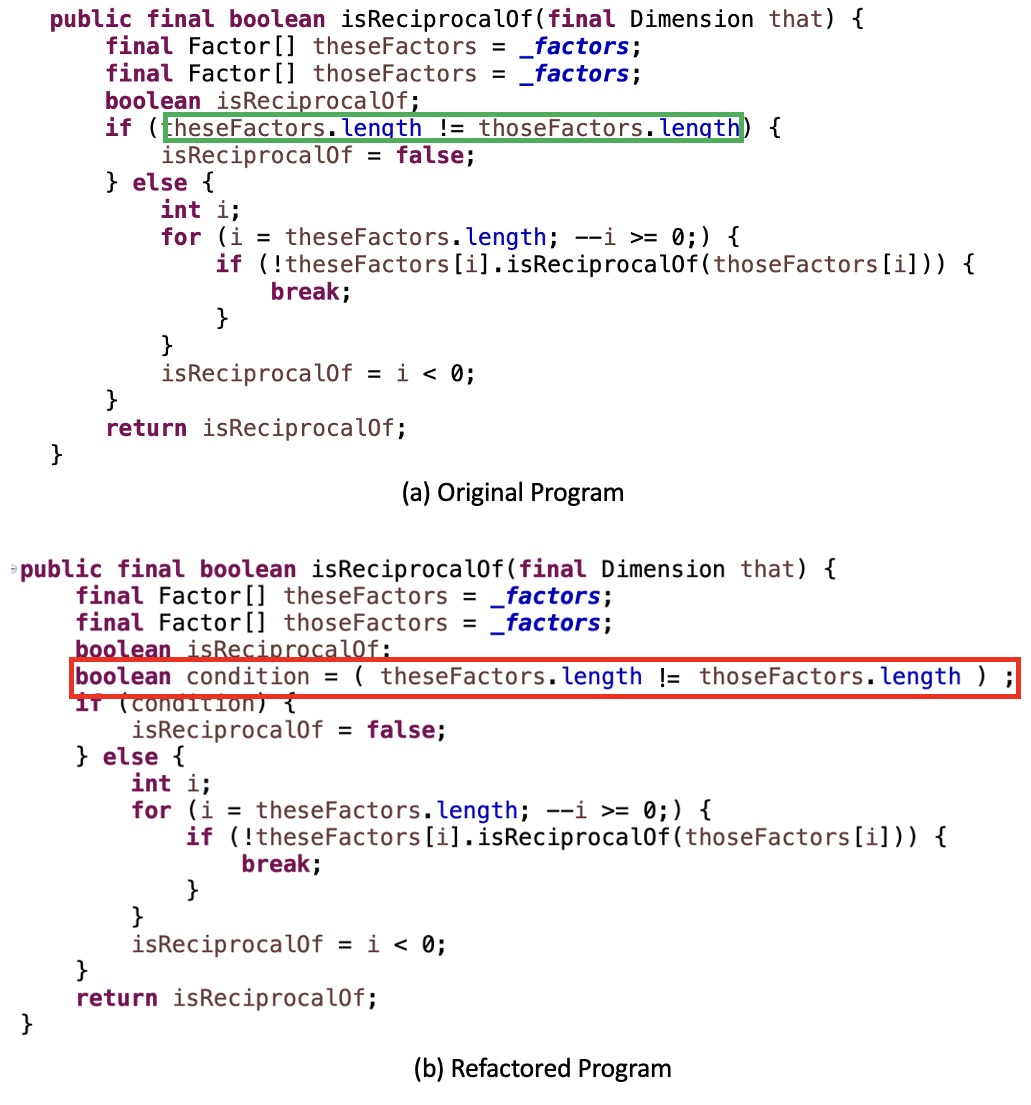}
    \caption{Impact of refactoring on the performance of GraphCodeBERT model in block swap transformation. Results show that the model's performance decreases by around 10\% with such operation in both original and transformed code, but is still quite good.}
    \vspace{-.6cm}
    \label{reposition}
\end{figure}

We consider a \emph{Refactoring} transformation, 
to see if this affects the performance of the model\footnote{We thank Aryaz Eghbali of the Uni. of Stuttgart, for this suggestion.}. 
Figure~\ref{reposition} presents an example showing how we refactored the conditional statement, by introducing a boolean variable, which was
assigned to hold the value of the condition in the conditional statement. Note that it is not trivial to change the position of the conditional statement; one has to ensure that  such a refactoring step leaves the semantics the same. We declared a boolean variable condition, assigned the conditional statement to the variable, and replaced the statement with that variable. After this refactoring step, we applied a block swap operation on the function and queried the model (on the original and block-swapped versions): the operator within right-hand-side of statement assigning the value of the boolean variable. In Figure~\ref{reposition}, \eg,  in the right hand side of the assignment
to the {\small\tt condition} variable, we mask out the ``!="  (unswapped) and ``==" in the (swapped) versions, and ask the model to recover the right operator.

\subsubsection{Distance in embedding space}
\label{sec:embedding}

We also study how program transformations affect
the representation space. Specifically, we create two versions of
programs from the original function. One preserves computational
meaning (i.e., block swap), and the other one is superficially more
similar (smaller edit distance) but is semantically non-equivalent
(See example in Figure~\ref{distance}). We take the embedding of the CLS token
in the last layer as the program representation and report the euclidean distance between the two swap versions and the original. The "CLS" token in a transformer model refers to a special token called the "classification token." It is typically added, as a marker, to the beginning of an input sequence;  the embedding thereof can be used to represent the entire input sequence, for tasks like text classification or sentence-level predictions. If PLMs  
 robustly capture meaning, in representation space,  \emph{semantically} identical programs should be closer in representation space, despite the non-equivalent programs being more similar \emph{in form}. Note that we don’t use any masks because it is not conclusively possible to infer the semantics with a version with a mask. On the other hand, for masking experiments, we don't need any input for the CLS token. The model’s API provides possible solutions for the masked position along with their associated probabilities.
We discuss the result in Section~\ref{dis-res}.

\section{Experiments and Results}
\input{swap_accuracy_results}

\subsection{Dataset, Models, and Experiments}
\label{data}
We implemented two transforms in our experiments to check the robustness property, \texttt{Block Swap}, and \texttt{Operand Swap}. For block swap, we parsed  programs into  abstract syntax tree (AST)\footnote{We use \url{https://github.com/tree-sitter/tree-sitter}} form, and swapped the ``then-else'' clause nodes in the tree; to preserve meaning, the conditional operator therein will have to  change. Similarly, for operand swap, we swap the variable names of the statement clause. We apply operand swap to all conditional statements (\emph{e.g.,} in both loops and conditionals), while we only apply transformation to if-else statements with block swap. 
We show examples in Figure~\ref{block} and~\ref{operand} for illustration. In all our examples, including these, we ensured that the transforms were meaning-preserving. 

We use Java code from the test set of CodeSearchNet~\cite{husain2019codesearchnet} for our experiments. 
CodeSearchNet is a properly de-duplicated dataset, using the approach due to Allamanis~\cite{allamanis2019adverse}. 
To mitigate data leakage from training to testing, we apply 
semantic transforms to \emph{just the test split} of the CodeSearchNet; 
the training split is used to pre-train at least one of the models used in our experiments. We apply several transformations (variable renaming and refactoring of the conditional statement) 
to the program; the resulting code is not what the original developer intended, 
and that also reduces the Likelihood of the exact test data being present in the original pre-training dataset (Casalnuovo \emph{et al}~\cite{casalnuovo2020programmers} reported that such ``unnatural" transformed code had lower log-likelihood, when measured with
well-trained models). 
We gather 1,425 examples for block swap and 9,377 examples for operand swap. To ensure the preservation of meaning, we manually reviewed all the samples. When employing the block swap transformation, we maintained the original order of the operands. This strategy makes it straightforward to guarantee meaning preservation for this transformation. However, when applying the operand swap transformation, altering the order of the operands can potentially modify the meaning of the expression (e.g., f1(a)>f2(b), f1(a)>b) because function calls may have side-effects. 
To avoid this risk, we conservatively 
discarded all cases where we could not ensure the absence of side-effects when evaluating the operands. We retained cases of the form A op B, where A and B are both variables, or constants, etc. We also retained examples where one of the operands is a function and another one is a constant (``f1(a)>5'' is acceptable but ``f1(a)>b'' is not acceptable, because f1(a) could have side effects). After this conservative selection, we were left with 9,126 examples. 

Specifically, we choose the programs where there are conditions that we can apply block or operand swap. 
For all the experiments, the task is to predict the masked operator (i.e., ==, !=, <, >, <=, >=).
We experiment with one LM trained on natural language (RoBERTa \cite{liu-etal-2019-roberta}) and two PLMs trained on code (CodeBERT~\cite{feng-etal-2020-codebert} and GraphCodeBERT~\cite{guo-etal-2021-graphcode}). In particular, CodeBERT adapts MLM and replaced token detection (RTD) as the training objective, while GraphCodeBERT also considers simplified dataflow (including edge prediction and node alignment) in addition to MLM. We report the model prediction accuracy before and after the transformation on Java, as well as the accuracy when the model predicts both forms correctly. A brief overview of the pre-trained models used in our experiments are given in Appendix.


\subsection{Results and Analysis}

\subsubsection{Accuracy of block swap and operand swap}
Table \ref{tab:accuracy} shows results for operator prediction accuracy. CodeBERT and GraphCodeBERT achieve high accuracy on the original programs ($> 80\%$) and relatively high for the transformed programs ($> 60\%$) for both the block swap and operand swap. 
We ran 1000 Monte-Carlo simulations, 
 where we randomly guessed the masked operators based on the prior frequency in the training set. None of the 1000 simulations yielded an accuracy value
 more than 38\%, suggesting that the observed value of $> 60\%$ is statistically significantly better than random guessing.

A more detailed analysis, also with a Monte-Carlo simulation
to get a p-value, is presented in \S ~\ref{sec:erroranalysis}.
The slightly lower accuracy\footnote{This lower accuracy is still
statistically significantly better than guessing randomly.} for transformed programs is unsurprising, given the often formulaic and idiomatic nature of code. Models naturally perform more accurately on text that is more familiar and has lower entropy\footnote{We only consider the operator for entropy calculation.}, making the original condition easier than the transformed condition. Still, it is important to remember that no instructions or explicit indication of what the code should do (separate from what is already in the code itself) is given to the model in either the original or transformed conditions. Additionally, in the transformed condition, the model is \emph{not} presented with what the original code was, and given generally unfamiliar code, must decide on an operator. Given that many operators would be syntactic valid in the same position, we conjecture that that the choice of operator must be driven by the model's representation of the semantics of the code, not its form, especially since in the transformed condition the form looks more unfamiliar in general. This suggests that 
the PLMs are robustly learning a realistic distribution over the \emph{semantics} of the code,
and can thus correctly perform the pre-training tasks, even  
when working with different forms of a program (while maintaining the same computational meaning). 

The greater accuracy of Operand Swap over Block Swap is noteworthy: our data suggests this maybe because a correct prediction in the case of Block Swap requires attention to a much longer sequence of subsequent text (see next section). 
The poor results of RoBERTa, which is a model of natural language and not pre-trained specifically on code data,
are noteworthy; the results suggest that 
it lacks robustness in capturing semantics of code, even though it appears to capture its syntax. This serves as a stronger baseline than one based solely on frequency counts. RoBERTa appears to have seen enough code to mimic its form to some extent, but not to capture its meaning.

\subsubsection{Impact of context length and direction}
\input{context_length}
In Section~\ref{clen}, we discuss the context length and direction used by the model for prediction.
Table \ref{tab:length} examines how  model prediction accuracy (and entropy) vary with context length. We evaluate the performance considering the next ten tokens only after the target operator (e.g., +10) or together with the previous ten tokens (e.g., $\pm$10). Results suggest that previous tokens  
(such as function names and identifiers)
help somewhat,  but most of the value is from longer contexts \underline{\emph{after}} the target token for both block swap and operand swap. Note that as the amount of context increases, the entropy decreases, indicating greater model confidence. For Block swap, using the 50 following tokens we achieve 82.05\% accuracy while adding 50 more preceding tokens increases the accuracy to 85.13\%. 
This suggests that the preceding tokens have 
a comparatively small impact on the models' accuracy. The importance of the following tokens to accuracy gives an additional signal that the model is utilizing the correct context
for determining the correct missing operator in both transformed and original programs. 

\subsubsection{Impact of consistent variable renaming}
\label{res_var_ren}
As discussed earlier in Section~\ref{cvr}, we also use consistent variable renaming to evaluate the models' accuracy and robustness against meaning preserving transformations.
The number of samples where the model successfully predicts the original and transformed operators decrease slightly (65.19\% vs. 64.64\% for block swap and 89.85\% vs. 86.11\% for operand swap); but both the performance and entropy of the model are overall similar to what we observed with the original version of the program. Therefore, variable renaming does not impact GraphCodeBERT's performance on both the original and transformed version of the code; This finding also suggests that the model is robustly 
capturing useful statistics of the \emph{semantics} of programs, even with variable renaming. 

For renaming, we choose unhelpful variable names, the kind that  professional developers have learned to avoid. The model is unlikely to have seen the transformed program in the pre-training dataset. Even so, the model fairly reliably predicts the operators, suggesting that the model robustly learns the semantics of the program and isn't just repeating something it has seen during pre-training.

\begin{table}[t]

\resizebox{.85\columnwidth}{!}{%
\renewcommand{\arraystretch}{1.2}

\centering
\begin{tabular}{llccccc}
\hline
\multicolumn{1}{c}{\multirow{2}{*}{Transformation}} & \multicolumn{1}{c}{\multirow{2}{*}{Model}} & \multicolumn{3}{c}{Accuracy}     & \multicolumn{2}{c}{Entropy} \\
\multicolumn{1}{c}{}                                & \multicolumn{1}{c}{}                       & Original & Transformed & Both    & Original    & Transformed   \\ \hline
\multirow{3}{*}{Bock Swap}                          & RoBERTa                                    & 50.04\%  & 27.4\%      & 1.87\%  & 1.48        & 1.81          \\
                                                    & CodeBERT                                   & 84.88\%  & 62.33\%     & 58.41\% & 0.56        & 1.41          \\
                                                    & GraphCodeBERT                              & 88.12\%  & 67.89\%     & 64.64\% & 0.46        & 1.39          \\ \hline
\multirow{3}{*}{Operand Swap}                       & RoBERTa                                    & 52.53\%  & 14.07\%     & 11.44\% & 0.38        & 1.45          \\
                                                    & CodeBERT                                   & 92.91\%  & 83.28\%     & 81.72\% & 0.26        & 0.71          \\
                                                    & GraphCodeBERT                              & 94.65\%  & 87.54\%     & 86.11\% & 0.20        & 0.55    \\ \hline     
\end{tabular}
}
\caption{Impact of variable renaming on the performance of the models in block swap transformation. Results show that the model's performance degrades by less than 1\%-4\%.}
\label{tab:obf}
\vspace{-.5cm}
\end{table}


\subsubsection{How do the models perform when condition expressions are refactored?}

\begin{table}[t]

\resizebox{.85\columnwidth}{!}{%
\renewcommand{\arraystretch}{1.2}

\centering
\begin{tabular}{lccccc}
\hline
\multicolumn{1}{c}{\multirow{2}{*}{Model}} & \multicolumn{3}{c}{Accuracy}   & \multicolumn{2}{c}{Entropy} \\
\multicolumn{1}{c}{}                       & Original & Transformed & Both  & Original    & Transformed   \\ \hline
RoBERTa (NLP)                              & 13.77\%    & 7.78\%        & 0.70\%   & 2.64        & 2.39          \\
CodeBERT-mlm                               & 68.52\%    & 51.06\%       & 41.83\% & 1.13        & 1.56          \\
GraphCodeBERT                              & 77.41\%    & 58.16\%       & 52.52\% & 0.88        & 1.45   \\ \hline       
\end{tabular}
}
\caption{Impact of refactoring conditional statement on model performance \emph{in re} block-swap. Performance decreases somewhat (by around 10\%) with refactoring, but is still quite good. }
\label{tab:repos}
\vspace{-.6cm}
\end{table}

We consider now the effect of  refactoring (See \autoref{sec:refactoring}) boolean expressions by introducing a boolean variable to replace the condition in the {\small\tt if} statement. 
Table~\ref{tab:repos}  shows that the model's accuracy decreases for all the models. For GraphCodeBERT, the accuracy goes down for both the original and block-swapped versions, and the model's accuracy to correctly guess both operators goes down from 65.19\% to 52.52\%. However, 52.52\% is significantly higher than random guessing. Similar to our earlier findings, none of the observed values in our 1000-run simulation exceeded 38\% percent. Note that refactoring the conditional statement, substantially increases the entropy of the operator, and (as shown below) the models' robustness depends on
this measure.

\input{embedding_similarity}

\subsubsection{Distance in embedding space}
\label{dis-res}
We also examined (see \S \ref{sec:embedding}) the geometric effect of program transformation in the vector representation space. Specifically, we create two versions of programs from the original function. One preserves computational meaning (\emph{i.e.,} block swap), and the other one is superficially more similar to the original program (block swap without changing the operator, which results in smaller edit distance) but is semantically non-equivalent (see example in Figure \ref{distance}). We take the embedding of the CLS token in the last layer as the program representation, and report the Euclidean distance between the two swap versions and the original. Table \ref{embedding-similarity} shows that the distance between semantics-preserving swap is significantly smaller than that for non-preserving swap. The results suggest that PLMs assign similar representations to programs of similar meanings, rather than similar surface forms. 
From the perspective of string similarity, the semantically non-equivalent program is strictly more similar to the original, but the other, less superficially similar program, has the same semantics.
Since the transformed programs do not occur in the training data and in fact are rather different from real code scripts, our findings suggest that code PLMs learn a robust representation of computational meaning, beyond superficial features.

\vspace{-.2cm}
\subsection{Error Analysis \& Entropy}
\label{sec:erroranalysis}



Table \ref{tab:error_analysis} (in Appendix) shows model performance on the original and block swapped programs for each individual operator. For example, for
the "!=" operator the "no" value in the ``block swapped" indicates that the 
model should correctly guess the original operators  and a ``yes" value indicates transformed operators. 
We observe that the model achieves the highest F-score for ``=='' ($0.94$); 
this conditional operator occurs most often in the original code.
However, after transformation, the model can still predict its corresponding ``!='' correctly ($0.82$) despite the  higher frequency distribution of ``=='' in the training data. Meanwhile, results show a large performance drop after the transformation for ``<'' and ``>'', where the model is confused by whether to add the equal condition (e.g., the model may predict ``>'' while the ground truth is ``>='' after the transformation for ``<''). 
We surmise that the lower performance on correctly predicting the "<=" and ">=" is due to the relatively low frequency in which they occur
in the original (untransformed) form: just 25 times for "<=" and 26 for  ">=" in {\small\tt if} statements where blocks could be swapped. 
This can also be justified by the relatively low F-score after transformation on the ``<='' and ``>='' operators, where the frequency distributions are relatively similar. We observed similar results with operand swap operations (see Table~\ref{tab:error_analysis1} in Appendix). Apart from  >= and <=, for all the operators the F-score is greater than 0.73. 
We conducted an evaluation of each operator using a Monte Carlo simulation, with 1000 separate runs.
We randomly assigned an operator based on the frequency distribution of the operators and compare it with the correct operators. 
In all scenarios, the distribution-driven random guessing under-performs the model, except for one case. Specifically, when considering the ``<='' operator, the model got an F-score of just 0.10 (in block swap before transformation). However, in only 15 out of 1000 runs (1.5\% of the time), the random assignment manged to outperform the model. Consequently, based on our statistical significance threshold ($p<0.01$) in all other scenarios (23 out of 24 considering both block and operand swap), the model demonstrates an ability to surpass the frequency distribution and effectively react to preserve the intended meaning of the code.

To examine whether the PLMs predict the operands by merely memorizing token frequency, rather than more robustly relying on the functional semantics (which is an important component of meaning), we  report the entropy (negative log-likelihood) of the masked operands. Entropy measures the unlikelihood of a predicted token, which is influenced by its surrounding context.
Table \ref{tab:accuracy} presents the log-likelihood of the original operator and transformed operator and the log-likelihood 
of the predicted operator increases significantly after the transformation (for example, doubling from $0.17$ to $0.35$ for GraphCodeBERT on Operand Swap). Even so the accuracy is robustly preserved (from $95.32\%$ to $92.14\%$). This suggests that rather than minimizing the superficial
(lexical) entropy by memorizing familiar syntax, PLMs learn to encode the computational meaning.

\subsection{How familiar are the perturbed samples?}
\label{sec:mem_syntax}
Using only the exact objective function (re-filling masked tokens) allows us to investigate pre-trained masked models. We believe this can be one of the methods to investigate a model's semantic understanding because the models were trained to unmask 15\% of randomly chosen tokens. This does not present a semantic understanding challenge, but rather a syntactic one. The question is whether having learned (perhaps ``memorized'') to fill in masked tokens in original text necessarily enables the model to perform the task with the perturbed samples. 

\vspace{.2cm}

\noindent\textbf{Finding exact conditional statements in the training corpus:}
Since we have access to the training and test samples, we can actually examine the conditional statements in the training dataset. We employed a brute-force search technique and observed that in 54.97\% of the samples, neither the original nor the transformed version is found in the training corpus (Table~\ref{tab:mem}), and in 35.37\% of the cases, both the original and transformed code are found in the corpus. The model’s accuracy is 84.90\% for the latter group, which is not surprising because the samples in that particular group may be used frequently, and the model performs well on that code. However, for the first group, where the model hasn’t seen any samples from the training set, achieving 53.11\% accuracy for both versions is quite impressive. That means for 53.11\% of the samples the model was able resolve semantic dilemmas even though it did not directly see those conditional statements. We also apply this technique to obfuscated code. As expected, a fraction of unseen samples go up to 82.67\% and the model still performs really well (even better) with 59.64\% both accuracy. However, overall both accuracy is lower for this group (64.64\% vs 65.19\%). Note that if we pick the majority operator it will result in syntactic correctness but both accuracy would be 0\%. Consider the scenario where the original conditional statement was found in the corpus but not the transformed one (row 3 in Table~\ref{tab:mem}). The accuracy for the original and transformed code is respectively 83.72\% and 47.67\%. Though the accuracy of the transformed code is low, it is still much higher, and we achieve an accuracy of 46.51\% for both. Additionally, the fraction of such samples is relatively low, only 7\%.

\vspace{.2cm}

\noindent\textbf{Model’s Familiarity with the target operator:}
We include a boxplot to show that the entropy is consistently higher for transformed code in the GraphCodeBERT model (see Figure~\ref{ent_gcbert} in Appendix). Although the average entropy (1.7) may not seem high, it is important to note that our goal of achieving meaning-preserving transformation limits how much higher the entropy of the transformed operator can be. While it might be possible to achieve higher entropy while still keeping the meaning intact, this could prevent us from evaluating the model at scale.
Answering the question of model familiarity is challenging from the masked model perspective due to the complete context of the program, which may help the model become more confident about the intended solution. However, from the perspective of an autoregressive model, it’s easier to determine which token the models prefer at a certain position. Using entropy scores from the Codegen-2 (7B parameter) model~\cite{nijkamp2023codegen2} (Figure~\ref{ent_codegen} in Appendix), we can still observe that the model prefers (is less ``surprised'' by) the original version of the code over the transformed one, even after refactoring and obfuscation. It's important to note that we are not claiming semantic understanding of the autoregressive model here; rather, we are demonstrating how the model is less “familiar” with the transformed code.

\begin{table}[t]

\resizebox{.90\columnwidth}{!}{%
\renewcommand{\arraystretch}{1.2}

\centering
\begin{tabular}{lcccccc}
\hline
\multicolumn{1}{c}{\multirow{2}{*}{Is it obfuscated?}}                                    & \multirow{2}{*}{\begin{tabular}[c]{@{}c@{}}Original\\ Found?\end{tabular}} & \multirow{2}{*}{\begin{tabular}[c]{@{}c@{}}Transformed \\ Found?\end{tabular}} & \multirow{2}{*}{\begin{tabular}[c]{@{}c@{}}Fraction (\%) \\ of Samples\end{tabular}} & \multicolumn{3}{c}{Accuracy in (\%)} \\
\multicolumn{1}{c}{}                                                                      &                                                                            &                                                                                &                                                                                      & Original   & Transformed  & Both     \\ \hline
\multirow{4}{*}{\begin{tabular}[c]{@{}l@{}}OriginalVariable\\ Names\end{tabular}}         & No                                                                         & No                                                                             & 54.97\%                                                                              & 84.28\%    & 56.72\%      & 53.11\%  \\
                                                                                          & No                                                                         & Yes                                                                            & 2.52\%                                                                               & 87.5\%     & 84.37\%      & 71.87\%  \\
                                                                                          & Yes                                                                        & No                                                                             & 7.22\%                                                                               & 83.72\%    & 47.67\%      & 46.51\%  \\
                                                                                          & Yes                                                                        & Yes                                                                            & 35.37\%                                                                              & 93.91\%    & 87.16\%      & 84.98\%  \\ \hline
\multirow{4}{*}{\begin{tabular}[c]{@{}l@{}}Uninformative\\ Variable\\ Names\end{tabular}} & No                                                                         & No                                                                             & 82.67\%                                                                              & 86.54\%    & 63.10\%      & 59.64\%  \\
                                                                                          & No                                                                         & Yes                                                                            & 5.33\%                                                                               & 100\%      & 97.29\%      & 97.29\%  \\
                                                                                          & Yes                                                                        & No                                                                             & 4.48\%                                                                               & 90.74\%    & 81.48\%      & 77.77\%  \\
                                                                                          & Yes                                                                        & Yes                                                                            & 7.53\%                                                                               & 94.04\%    & 86.90\%      & 84.52\%  \\ \hline
\end{tabular}
}
\caption{Fraction of samples found/not-found in the training corpus both for the normal and obfuscated code. We also present GraphCodeBERT’s accuracy on original code, transformed code and both (where the model successfully predicts operators in the original and transformed versions).
}
\label{tab:mem}
\vspace{-1cm}
\end{table}

\section{Limitations}
Our experiments and analysis suggest that PLMs can learn a  semantic representation that is robust against some meaning-preserving transforms; but  these findings may only apply to code PLMs rather than PLMs for natural language. Furthermore, the semantics of code is quite different from Natural Language semantics, although parallels exist. 
Since defining ``meaning understanding'' is complicated in natural language, it would be non-trivial to construct an experiment like ours, in natural language. 
Nevertheless, we hope to expand our experiments to natural language inspired by pragmatics and psycho-linguistic diagnostics \cite{parrish-etal-2021-nope, ettinger-2020-bert} as our future work.
Also, our proposed approach can only investigate the masked language model. The recent large language models (\eg GPT-3) are decoder only, and whether that auto-regressive generative model learns semantics cannot be judged by our approach; furthermore  evaluating such models (whose training data is so
vast, and/or opaque) is complicated by the challenges
of ensuring training/test separation. 
Nevertheless, using a linear probing in program synthesis, Jin and Rinard successfully extracted abstractions of both current and future program states from the model states in the auto-regressive model. This achievement suggests that the model possesses an understanding of semantics~\cite{jin2023evidence}.

\section{Conclusion}
Unlike previous probing research using linear classifiers, we study how much pre-trained language models encode semantics for understanding beyond frequency and co-occurrence. We designed experiments in a restricted setting based on the precise and formal quality of code and found that the models learn code semantics. We also observed that semantically equivalent program pairs have the minimum distance in the embedding space compared to syntactically more similar but semantically different program pairs. Our experiments with restricted context, variable renaming, and repositioning the conditional statement strengthen our claim about the model's semantic understanding because such transformations made the program more unnatural. Still, the model can efficiently predict the original and transformed operators correctly even though the model is not pre-trained with the exact token sequence.

\balance
\bibliography{custom}
\bibliographystyle{acm}

\newpage

\appendix

\section{Appendix A}

A brief overview of the pre-trained models used in our experiments are given below.
\subsection{RoBERTa}
BERT~\cite{devlin-etal-2019-bert} was an early transformer model that used pre-training as a strategy; it surpassed earlier transformer models in a variety of NLP tasks. It used two pre-training objectives: Masked Language Modeling (MLM) and Next Sentence Prediction (NSP). MLM  randomly masks out 15\% of the tokens and trains the model to predict them; NSP trains the model to predict the next sentence following an input sentence. RoBERTa, proposed by Liu et al.~\cite{liu-etal-2019-roberta}, improved on BERT's performance by implementing a few changes, such as dynamic masking and dropping NSP. As a result, it achieved better results and is used as the NLP baseline model.
RoBERTa was trained with 160GB of uncompressed text from five English-language corpora, including Bookcorpus, CC-News, Openwebtext, and Stories.

\subsection{CodeBERT}
CodeBERT~\cite{feng-etal-2020-codebert} is similar in structure to the RoBERTa model and utilizes two pre-training objectives: MLM and Replaced Token Detection (RTD). RTD involves two data generators, NL and PL, generating possible replacements for a set of randomly masked positions, which the model is then trained to classify as either the original word or a replacement. CodeBERT was pre-trained on the CodeSearchNet dataset. Note that we could not use the original CodeBERT model because that model is not easily programmable. We used an alternative CodeBERT version ``CodeBERT-mlm'', pre-trained with only MLM objective.
\subsection{GraphCodeBERT}
GraphCodeBERT~\cite{guo-etal-2021-graphcode} augments source code with data-flow, during  pre-training. It uses a data flow graph (DFG); the DFG nodes are variable occurrences, while the edges represent value flow. GraphCodeBERT is pre-trained with three objectives (Edge Prediction, Node Alignment, and MLM) on 2.3 million functions (PL-NL pairs) from the CodeSearchNet dataset. It learns a joint representation of the DFG structure, DFG alignment with source code, and the source code token sequences. This approach utilizes high-capacity models that are pre-trained over a large, multilingual corpus. Hence, the models already have extensive knowledge of each language even before fine-tuning.
Both CodeBERT and GraphCodeBERT were trained with the CodeSearchNet dataset, which includes 2.1 million bimodal data points and 6.4 million unimodal codes across six programming languages: Python, Java, JavaScript, PHP, Ruby, and Go.

\begin{figure*}[h!]
    \centering
    \includegraphics[scale=0.45, trim={0cm 0cm 0cm 0cm}]{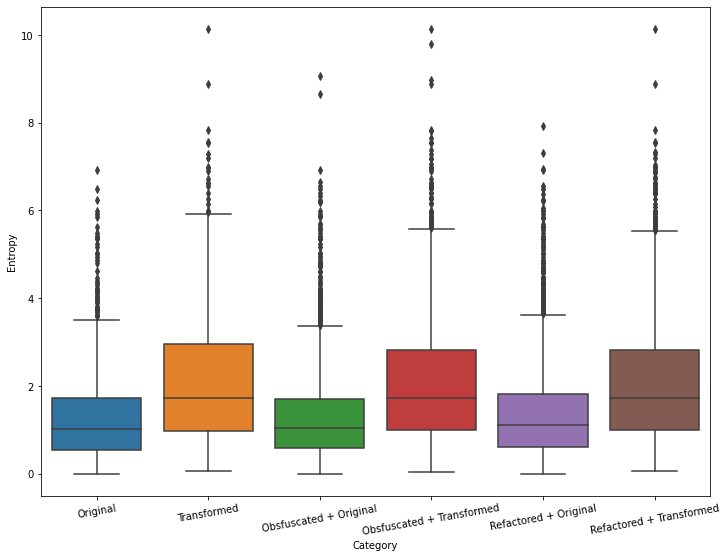}
    \caption{Entropy of the operator based on Codegen2 model}
    \label{ent_codegen}
\end{figure*}

\begin{figure*}[h!]
    \centering
    \includegraphics[scale=0.45, trim={0cm 0cm 0cm 0cm}]{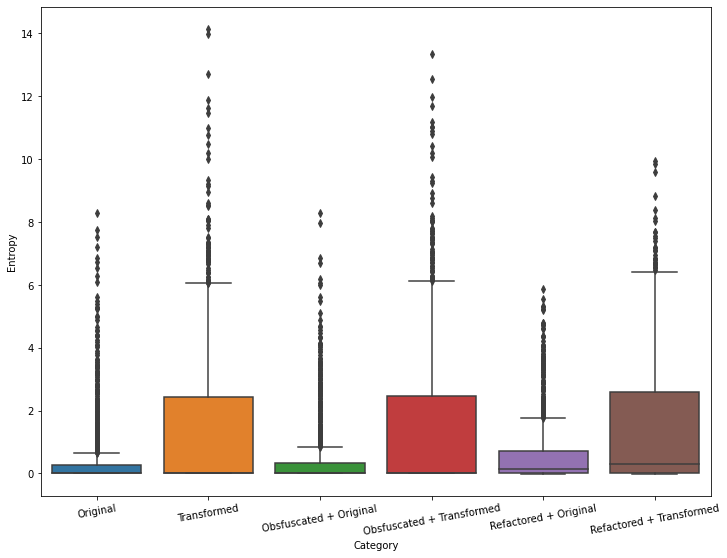}
    \caption{Entropy of the operator based on GraphCodeBERT model}
    \label{ent_gcbert}
\end{figure*}

\input{error_analysis_table}

\end{document}

%% file: swap_accuracy_results.tex
\begin{table}[t]

\resizebox{.90\columnwidth}{!}{%
\renewcommand{\arraystretch}{1.2}

\centering
\begin{tabular}{llccccc}
\hline
\multicolumn{1}{c}{\multirow{2}{*}{Transformation}} & \multicolumn{1}{c}{\multirow{2}{*}{Model}} & \multicolumn{3}{c}{Accuracy}     & \multicolumn{2}{c}{Entropy} \\ 
\multicolumn{1}{c}{}                                & \multicolumn{1}{c}{}                       & Original & Transformed & Both    & Original    & Transformed   \\ \hline
\multirow{3}{*}{Block Swap}                         & RoBERTa (NLP)                              & 48.12\%  & 29.01\%     & 1.88\%  & 1.43        & 1.78          \\
                                                    & CodeBERT                              & 84.64\%  & 63.65\%     & 59.04\% & 0.55        & 1.43          \\
                                                    & GraphCodeBERT                              & 87.97\%  & 68.34\%     & 65.19\% & 0.45        & 1.39          \\
\hline
\multirow{3}{*}{Operand Swap}                       & RoBERTa (NLP)                              & 52.57\%  & 12.95\%     & 10.40\% & 1.21        & 1.73          \\
                                                    & CodeBERT                               & 93.47\%  & 87.97\%     & 86.83\% & 0.24        & 0.48          \\
                                                    & GraphCodeBERT                              & 95.32\%  & 92.14\%     & 90.66\% & 0.17        & 0.35  \\ \hline        
\end{tabular}

}
\caption{Performance of model prediction on the original and transformed programs. Results show that the PLMs can still predict the operators accurately on both block and operand swap despite higher entropy. This indicates that the model relies on semantic meaning rather than simple frequency (signified by entropy).}
\label{tab:accuracy}
\vspace{-.9cm}
\end{table}

%% file: context_length.tex
\begin{table}[t]

\resizebox{.85\columnwidth}{!}{%
\renewcommand{\arraystretch}{1.2}

\centering
\begin{tabular}{llccccc}
\hline
\multicolumn{1}{c}{Transformation} & \multicolumn{1}{c}{Context} & \multicolumn{1}{c}{Original} & \multicolumn{1}{c}{Transformed} & \multicolumn{1}{c}{Both} & \multicolumn{1}{c}{\begin{tabular}[c]{@{}c@{}}Original \\ (Entropy)\end{tabular}} & \multicolumn{1}{c}{\begin{tabular}[c]{@{}c@{}}Transformed \\ (Entropy)\end{tabular}} \\ \hline
\multirow{7}{*}{Block-swap}          & $\pm$10                         & 66.61\%                       & 47.96\%                          & 31.35\%                   & 1.10                                                                               & 1.85                                                                                  \\
                                     & +10                          & 67.25\%                       & 46.14\%                          & 30.71\%                   & 1.18                                                                               & 2.03                                                                                  \\
                                     & $\pm$30                         & 81.77\%                       & 60.66\%                          & 52.80\%                   & 0.66                                                                               & 1.52                                                                                  \\
                                     & +30                          & 79.52\%                       & 58.34\%                          & 50.07\%                   & 0.77                                                                               & 1.58                                                                                  \\
                                     & $\pm$50                         & 85.13\%                       & 64.10\%                          & 58.34\%                   & 0.53                                                                               & 1.47                                                                                  \\
                                     & +50                          & 82.05\%                       & 61.99\%                          & 54.63\%                   & 0.66                                                                               & 1.54                                                                                  \\
                                     & Complete                     & 87.97\%                       & 68.34\%                          & 65.19\%                   & 0.45                                                                               & 1.39                                                                                  \\ \hline
\multirow{7}{*}{Operand-swap}        & $\pm$10                         & 85.05\%                       & 81.70\%                          & 78.08\%                   & 0.51                                                                               & 0.65                                                                                  \\
                                     & +10                          & 78.88\%                       & 73.83\%                          & 69.32\%                   & 0.69                                                                               & 0.98                                                                                  \\
                                     & $\pm$30                         & 92.86\%                       & 89.49\%                          & 87.46\%                   & 0.25                                                                               & 0.41                                                                                  \\
                                     & +30                          & 88.19\%                       & 84.39\%                          & 80.47\%                   & 0.38                                                                               & 0.57                                                                                  \\
                                     & $\pm$50                         & 93.93\%                       & 90.14\%                          & 88.44\%                   & 0.21                                                                               & 0.39                                                                                  \\
                                     & +50                          & 89.21\%                       & 85.60\%                          & 81.89\%                   & 0.36                                                                               & 0.54                                                                                  \\
                                     & Complete                     & 95.32\%                       & 92.14\%                          & 90.66\%                   & 0.17                                                                               & 0.35    \\ \hline                                                                              
\end{tabular}
}
\caption{Impact of context length and direction on the performance of GraphCodeBERT in block swap and operand swap transformations. Results show that the next few tokens (e.g., +50) are critical to model performance, while considering previous tokens (e.g., +-50) provides complementary information.
``complete'' considers all tokens in the program.}
\label{tab:length}
\vspace{-.8cm}
\end{table}

%% file: embedding_similarity.tex
\begin{table}[t]

\resizebox{.95\columnwidth}{!}{%
\renewcommand{\arraystretch}{1.2}

\centering
\begin{tabular}{lccc}

\hline

\multicolumn{1}{c}{Model} & \begin{tabular}[c]{@{}c@{}}distance between \\non-equivalent swap \end{tabular} & \begin{tabular}[c]{@{}c@{}}distance between \\equivalent swap\end{tabular} & p-value          \\ \hline
CodeBERT                  & 2.51e-5                                                         & \textbf{1.80e-5}                                                         & \textless{}0.001 \\
GraphCodeBERT             & 9.09e-5                                                         & \textbf{7.63e-5}                                                         & \textless{}0.001 \\ \hline
\end{tabular}
}
\caption{The impact of semantically meaning preserving transformations in the embedding space reported by the averaged Euclidean distance between semantically equivalent swap and non-equivalent swap. The significant difference suggests that PLMs learn a robust representation beyond superficial features. p-value is calculated for one-sided pairwise Wilcoxon test \cite{wilcoxon1992individual}.}
\label{embedding-similarity}
\vspace{-.8cm}
\end{table}

%% file: error_analysis_table.tex
\begin{table}[b]

\resizebox{.85\columnwidth}{!}{%
\renewcommand{\arraystretch}{1.2}

\centering
\begin{tabular}{llcccccc}
\hline
\multicolumn{1}{c}{Operator}       & \multicolumn{1}{c}{\begin{tabular}[c]{@{}c@{}}Is block \\ swapped?\end{tabular}} & TP  & FP  & FN  & Precision & Recall & F-score \\ \hline
\multirow{2}{*}{"=="}              & No                                                                               & 547 & 50  & 19  & 0.92      & 0.97   & 0.94    \\
                                   & Yes                                                                              & 314 & 171 & 7   & 0.65      & 0.98   & 0.78    \\
\multirow{2}{*}{"!="}              & No                                                                               & 301 & 15  & 20  & 0.95      & 0.94   & 0.94    \\
                                   & Yes                                                                              & 431 & 13  & 175 & 0.97      & 0.71   & 0.82    \\
\multirow{2}{*}{"\textless{}"}     & No                                                                               & 66  & 23  & 29  & 0.74      & 0.69   & 0.71    \\
                                   & Yes                                                                              & 18  & 67  & 8   & 0.21      & 0.69   & 0.32    \\
\multirow{2}{*}{"\textless{}="}    & No                                                                               & 2   & 11  & 23  & 0.15      & 0.08   & 0.1     \\
                                   & Yes                                                                              & 9   & 2   & 90  & 0.82      & 0.09   & 0.16    \\
\multirow{2}{*}{"\textgreater{}"}  & No                                                                               & 57  & 25  & 42  & 0.7       & 0.58   & 0.63    \\
                                   & Yes                                                                              & 9   & 107 & 16  & 0.08      & 0.36   & 0.13    \\
\multirow{2}{*}{"\textgreater{}="} & No                                                                               & 18  & 16  & 8   & 0.53      & 0.69   & 0.6     \\
                                   & Yes                                                                              & 20  & 10  & 75  & 0.67      & 0.21   & 0.32    \\ \hline
\multirow{2}{*}{Overall}           & No                                                                               & 991 & 140 & 141 & 0.88      & 0.88   & 0.88    \\
                                   & Yes                                                                              & 801 & 370 & 371 & 0.68      & 0.68   & 0.68  \\ \hline  
\end{tabular}
}
\caption{Operator-wise performance of GraphCodeBERT in the block-swapped program.}
\label{tab:error_analysis}
\vspace{-1em}
\end{table}

\begin{table}[b]

\resizebox{.85\columnwidth}{!}{%
\renewcommand{\arraystretch}{1.2}

\centering
\begin{tabular}{llcccccc}
\hline
\multicolumn{1}{c}{Operator}                 & \multicolumn{1}{c}{\begin{tabular}[c]{@{}c@{}}Is operand \\ swapped?\end{tabular}} & \multicolumn{1}{c}{TP}   & \multicolumn{1}{c}{FP}  & \multicolumn{1}{c}{FN}  & \multicolumn{1}{c}{Precision} & \multicolumn{1}{c}{Recall} & \multicolumn{1}{c}{F-score} \\ \hline
\multirow{2}{*}{"=="}                          & No                                                                                  & 4081                      & 90                       & 116                      & 0.98                           & 0.97                        & 0.97                         \\
                                               & Yes                                                                                 & 4119                      & 226                      & 78                       & 0.95                           & 0.98                        & 0.96                         \\
\multirow{2}{*}{"!="}                          & No                                                                                  & 2910                      & 62                       & 74                       & 0.98                           & 0.98                        & 0.98                         \\
                                               & Yes                                                                                 & 2901                      & 94                       & 83                       & 0.97                           & 0.97                        & 0.97                         \\
\multirow{2}{*}{"\textless{}"}                 & No                                                                                  & 761                       & 80                       & 67                       & 0.9                            & 0.92                        & 0.91                         \\
                                               & Yes                                                                                 & 184                       & 207                      & 125                      & 0.85                           & 0.67                        & 0.75                         \\
\multirow{2}{*}{"\textless{}="}                & No                                                                                  & 51                        & 46                       & 37                       & 0.53                           & 0.58                        & 0.55                         \\
                                               & Yes                                                                                 & 70                        & 65                       & 104                      & 0.52                           & 0.4                         & 0.45                         \\
\multirow{2}{*}{"\textgreater{}"}              & No                                                                                  & 237                       & 86                       & 73                       & 0.73                           & 0.76                        & 0.74                         \\
                                               & Yes                                                                                 & 558                       & 97                       & 270                      & 0.85                           & 0.67                        & 0.75                         \\
\multirow{2}{*}{"\textgreater{}="}             & No                                                                                  & 126                       & 46                       & 48                       & 0.73                           & 0.72                        & 0.72                         \\
                                               & Yes                                                                                 & 10                        & 25                       & 78                       & 0.29                           & 0.11                        & 0.16                         \\ \hline
\multicolumn{1}{l}{\multirow{2}{*}{Overall}} & \multicolumn{1}{l}{No}                                                             & \multicolumn{1}{c}{8166} & \multicolumn{1}{c}{410} & \multicolumn{1}{c}{415} & \multicolumn{1}{c}{0.95}      & \multicolumn{1}{c}{0.95}   & \multicolumn{1}{c}{0.95}    \\ 
\multicolumn{1}{l}{}                         & \multicolumn{1}{l}{Yes}                                                            & \multicolumn{1}{c}{7842} & \multicolumn{1}{c}{714} & \multicolumn{1}{c}{738} & \multicolumn{1}{c}{0.92}      & \multicolumn{1}{c}{0.91}   & \multicolumn{1}{c}{0.91}    \\ \hline
\end{tabular}
}
\caption{Operator-wise performance of GraphCodeBERT in the operand-swapped program.}
\label{tab:error_analysis1}
\vspace{-1em}
\end{table}